\documentstyle[amsfonts,prd,aps,epsf]{revtex}

\begin{document}
\title{Slicing and Brane Dependence of the (A)dS/CFT Correspondence}
\author{Amir Masoud Ghezelbash$^{(a,c)}$, Daisuke Ida$^{(b)}$, Robert B. Mann$^{(a)}$
and Tetsuya Shiromizu$^{(b)}$}
\address{$^{(a)}$Department of Physics, University of Waterloo, Waterloo,
 Ontario N2L 3G1, Canada\\
}
\address{$^{(b)}$Research Center for the Early Universe (RESCEU), The University of Tokyo,
 Tokyo 113-0033, Japan\\
}
\address{$^{(c)}$Department of Physics, Alzahra University, Tehran 19834, Iran\\
}
\maketitle

\begin{abstract}
We investigate the slicing dependence of the relationship between conserved
quantities in the (A)dS/CFT correspondence. Specifically, we show that the
Casimir energy depends upon the topology and geometry of spacetime
foliations of the bulk near the conformal boundary. We point out that the
determination of the brane location in brane-world scenarios exhibits a
similar slicing dependence, and we comment on this in the context of the
AdS/CFT correspondence conjecture.
\end{abstract}


\vskip1cm


\section{Introduction}

The AdS/CFT correspondence conjecture is now a fundamental concept in a
variety of subjects\cite{AdSCFT}. It is intimately connected with the
holographic renormalization group\cite{RG} and emerges in brane-world
scenarios \cite{Misao,Lisa2,Ida}. An extension of the conjecture to
asymptotically de Sitter spacetimes (the dS/CFT correspondence) has recently
been proposed by Strominger\cite{dSCFT}. While it can in certain ways be
trivially achieved by just Wick rotation of appropriate AdS/CFT quantities,
there are still many unclear points as to its formulation and
interpretation. However since the dS/CFT correspondence offers the hope of
providing an answer for the origin of the vacuum energy or the deSitter
entropy\cite{Area}, it remains a subject of lively investigation\cite
{deSitter,SIT,Mann,State}.

Supporting evidence for the AdS/CFT conjecture has been provided from the
computation of conserved quantities. \ Specifically, it has been shown that
the usage of CFT-inspired boundary counterterms furnishes a means for
calculating gravitational actions and conserved quantities without reliance
on any \ reference spacetime \cite{CFT}. \ Moreover, calculations of the
total energy at infinity in odd-dimensional spacetimes yield additional
contributions to the total energy of black hole spacetime that could be
interpreted as the Casimir energy of the boundary CFT. \ These results have
recently been extended to asymptotically de Sitter spacetimes with success 
\cite{Mann}.

\ In this paper we consider the slice(brane) dependence of conserved
quantities\footnote{%
An early discussion of the dependence\ of the Casimir energy on boundary
geometry appeared in ref. \cite{Skenderis}.}. Specifically, we demonstrate
that the additional Casimir-type contributions to the total energy are
dependent on the slicing topology, as one might expect from quantum field
theory in curved spacetimes\cite{QFT}. We also find a dependence on the
choice of slicing geometry; this issue is presumably related to the recent
discussion of the state of the CFT \cite{State}. We then consider a similar
situation in brane-world scenarios \cite{Misao,Lisa2,Ida}, examining a
recent claim by Verlinde that the Friedman equation on the brane is
identical to the Cardy formula\cite{Cardy} if the correspondence is such
that the curvature term is identified with the Casimir energy \cite{linde}.
Following the standard AdS/CFT correspondence, on the other hand, we can
derive the effective gravitational equation on the curved branes and point
out a gap between Verlinde's argument and the standard one.

\section{(A)dS/CFT correspondence and slicing}

The energy-momentum tensor of the boundary CFT as determined by the
counter-term method is \cite{CFT} 
\begin{eqnarray}
T_{\mu \nu }^{\text{CFT}} &=&\frac{1}{8\pi G_{n}}\biggl[K_{\mu \nu }-K\gamma
_{\mu \nu }  \nonumber \\
&&~~~~~~~~-\frac{1}{\ell }\sum_{k=1}^{\left[ n/2\right] }\Theta \left(
n+1-2k\right) \ell ^{2k}T_{\mu \nu }^{(k)}\biggr]  \label{TCFT1} \\
&=&\frac{1}{8\pi G_{n}}\biggl[K_{\mu \nu }-K\gamma _{\mu \nu }-\frac{\left(
n-2\right) \Theta \left( n-1\right) }{\ell }\gamma _{\mu \nu }  \nonumber \\
&&~~~~~~~~+\frac{\ell \Theta \left( n-3\right) }{(n-3)}^{(n-1)}G_{\mu \nu
}+\cdots \biggr],  \label{TCFT2}
\end{eqnarray}
where $\gamma _{\mu \nu }$ is the metric for a given choice of boundary
slice ${\cal B}$ of spacetime ${\cal M},$ whose extrinsic curvature is $%
K_{\mu \nu }$ and whose unit normal is $u^{\mu }$. \ The step function $%
\Theta (x)$ vanishes for $x\leq 0$, and so the terms in the ellipsis are
non-zero only for $n\geq 6$. We have explicitly written the first two terms
in the sum, where $^{(n-1)}G_{\mu \nu }$ is the $(n-1)$-dimensional Einstein
tensor of the boundary. Every term is a rank-2 tensor consisting of terms
proportional to powers and/or derivatives of the boundary curvature; they
have been explicitly computed up to $n=9$ \cite{Mann,Das}. \ 

Given a Killing vector $\xi ^{\nu }$, the quantity 
\begin{equation}
{\frak Q}_{\xi }=\oint_{\Sigma }d^{n-2}\varphi \sqrt{\sigma }n^{\mu }T_{\mu
\nu }^{\text{CFT}}\xi ^{\nu }  \label{Qcons}
\end{equation}
is defined as the conserved charge associated with an $(n-2)$-dimensional
surface $\Sigma \subset {\cal B}$, whose unit normal $n^{\mu }$ is
orthogonal to $u^{\mu }$ ($u\cdot n=0$). \ \ 

\subsection{Static Slicings}

As an example, consider an evaluation of the energy of Schwarzschild-AdS
spacetime with static slicing, that is, 
\begin{equation}
ds^{2}=-f(r)dt^{2}+\frac{1}{f(r)}dr^{2}+r^{2}\sigma _{ij}dx^{i}dx^{j},
\label{SADS}
\end{equation}
where $\sigma _{ij}$ is a metric of a unit $\left( n-2\right) $-dimensional
sphere, plane or (compact)\ hyperboloid for $k=1,0,-1$, respectively, and $
%
f(r)=k-2m/r^{n-3}+r^{2}/\ell ^{2}$. In this slice the energy is measured at $%
r=\infty $. The extrinsic curvature of a surface with spacelike unit normal $%
u^{\alpha }=\hat{r}^{\alpha }=\left( \partial /\partial r\right) ^{\alpha }$
is 
\begin{equation}
K_{tt}=-\frac{f^{\prime }}{2}\sqrt{f(r)}~~~~~~~K_{ij}=r\sqrt{f(r)}\sigma
_{ij}  \label{EXSADS}
\end{equation}
and 
\begin{eqnarray}
T_{tt} &=&\frac{\left( n-2\right) f}{8\pi G_{n}\ell }\biggl\{-\frac{\ell 
\sqrt{f}}{r}+\left[ \Theta \left( n-1\right) +\frac{k\ell ^{2}\Theta \left(
n-3\right) }{2r^{2}}\right]  \nonumber  \label{stress3} \\
&&-\frac{k^{2}\ell ^{4}\Theta \left( n-5\right) }{8r^{4}}+\frac{k^{3}\ell
^{6}\Theta \left( n-7\right) }{16r^{6}}\biggr\}
\end{eqnarray}
where we have explicitly computed all terms up to $n=9$.

Consequently the total energy at $r=\infty $ is 
\begin{eqnarray}
{\frak M}_{{\rm AdS}}^{\left( n,k\right) } &=&\int_{\sigma }d^{n-2}x\sqrt{%
\sigma }T_{\mu \nu }^{\text{CFT}}n^{\mu }\xi ^{\nu }=V^{n-2}\frac{\left(
n-2\right) }{8\pi G_{n}}  \label{MSADS} \\
&&\times \left[ m+\frac{\Gamma \left( \frac{2p-3}{2}\right) \ell
^{2p-4}\left( -k\right) ^{(p-1)}}{2\sqrt{\pi }\Gamma \left( p\right) }\delta
_{2p-1,n}\right]  \nonumber
\end{eqnarray}
where $\xi ^{\nu }$ is the timelike Killing vector $\partial /\partial t$
and $V^{n-2}$ is the volume of the compact $\left( n-2\right) $-dimensional
space (for $k=1$, see \cite{Das}). If $p=\frac{n+1}{2}$ is a positive
integer, the second term remains. We see here the explicit dependence of the
CFT energy on the slicing topology. With spherical (hyperbolic) slicing, $%
k=1~(-1)$, the $r={\rm constant}$ hypersurface has ${\Bbb R\times }%
S^{n-2}(H^{n-2})$ topology and the Casimir energy naturally emerges; its
sign depends on the dimensionality and the value of $k$. However flat slices
with $k=0$ have ${\Bbb R}^{4}$ topology, and the Casimir energy vanishes.
This is of course compatible with the features of quantum field theory in
the curved spacetimes.

Similar results hold for the Schwarzschild de Sitter case. Working outside
of the cosmological horizon, the metric is \ 
\begin{equation}
ds^{2}=-\tilde{f}(r)dr^{2}+\frac{dt^{2}}{\tilde{f}(r)}+r^{2}d\tilde{\Omega}%
_{d-1}^{2}  \label{Sdsmet2}
\end{equation}
where 
\begin{equation}
\tilde{f}(r)=\left( \frac{r^{2}}{\ell ^{2}}+\frac{2m}{r^{d-2}}-1\right) ^{-1}
\label{flapse}
\end{equation}
Now the roles of the unit normals are inverted: the timelike unit normal is $%
u^{\alpha }=\widehat{t}^{\alpha }=\left( \partial /\partial t\right)
^{\alpha }$ , and a calculation analogous to the one above yields \cite{Mann}
\begin{equation}
{\frak M}_{{\rm dS}}^{\left( n\right) }=V^{n-2}\frac{\left( n-2\right) }{%
4\pi G_{n}}\left[ -m+\frac{\Gamma \left( \frac{2p-3}{2}\right) \ell ^{2p-4}}{%
2\sqrt{\pi }\Gamma \left( p\right) }\delta _{2p-1,n}\right]  \label{MSDS}
\end{equation}
where $\xi =\partial /\partial t$ is now a spacelike Killing vector, and the
overall sign-flip in the mass parameter arises from the relative signature
change in the boundary.

\subsection{Conformally Asymptotically Flat Slicings}

There is also a dependence of the CFT energy on the slicing geometry. \
Consider a slicing such that the boundary extrinsic curvature is given by 
\begin{equation}
K_{\mu \nu }=-\frac{1}{\ell }\gamma _{\mu \nu }  \label{Kmax}
\end{equation}
which we refer to as the conformally asymptotically flat (CAF) slicing
condition. For the exact $n-$dimensional deSitter spacetime, the induced
geometry on the slices is $(n-1)$-dimensional Euclid space. In this case the
first three terms in the CFT boundary stress-energy exactly cancel, yielding 
\begin{equation}
T_{\mu \nu }^{\text{CFT}}=\frac{1}{8\pi G_{n}}\frac{\ell \Theta \left(
n-3\right) }{(n-3)}\,^{(n-1)}G_{\mu \nu }+\cdots  \label{TCFTmax}
\end{equation}
Since the spacetimes under consideration are asymptotically (A)dS, \ we
suppose further that 
\begin{equation}
\gamma _{\mu \nu }=\Omega ^{2}\tilde{\gamma}_{\mu \nu }  \label{GAM}
\end{equation}
where the induced metric $\widetilde{\sigma }_{\mu \nu }$ on the
hypersurface $\Sigma $ is obtained from $\tilde{\gamma}_{\mu \nu }$, which
is asymptotically flat, and $\Omega ^{-1}$ vanishes on the conformal
completion of the boundary manifold. \ \ \ It is then straightforward to
show on dimensional grounds that 
\begin{equation}
T_{\mu \nu }^{(k)}\left( \gamma \right) =\Omega ^{2(1-k)}\tilde{T}_{\mu \nu
}^{(k)}\left( \tilde{\gamma}\right)  \label{ST}
\end{equation}
for every term in the sum (\ref{TCFT1}). \ Consequently, we have 
\begin{equation}
{\frak Q}_{\xi }=\frac{1}{8\pi G_{n}}\frac{\ell \Theta \left( n-3\right) }{%
(n-3)}\oint_{\Sigma _{\infty }}d^{n-2}\varphi \sqrt{\sigma }^{(n-1)}G_{\mu
\nu }n^{\mu }\xi ^{\nu }  \label{CC}
\end{equation}
as the surfaces $\Sigma $ approach the conformal completion $\Sigma _{\infty
}$ of the boundary. Note that this is an exact result, valid for $n>3$.

Now we can evaluate ${\frak Q}_{\xi }$ of the Schwarzshild-dS spacetime in
the CAF slicing. The metric has the form\cite{SIT} 
\begin{equation}
ds^{2}=-g^{2}d\tau ^{2}+a^{2}\omega ^{4/(n-3)}(d\vec{x}\cdot d\vec{x}),
\label{sitmet}
\end{equation}
where $d\vec{x}\cdot d\vec{x}$ is the metric of the $(n-1)$-dimensional
Euclidean space, 
\begin{equation}
\omega (\tau ,\vec{x})=1+\frac{m}{2(a(\tau )|\vec{x}|)^{n-3}}\text{ \ \ \ }%
g(\tau ,\vec{x})=\frac{2}{\omega }-1  \label{gomega}
\end{equation}
and $a(\tau )=e^{-\tau /\ell }$.

In this case $\xi =\partial /\partial t -(x^i/\ell)(\partial/\partial x^i)$%
,\ the CFT stress-energy is given by 
\begin{equation}
T_{\mu \nu }^{\text{CFT}}=-\frac{1}{8\pi G_{n}}\left( \delta _{ij}-\frac{%
\left( n-1\right) x^{i}x^{j}}{\left| \vec{x}\right| ^{2}}\right) \frac{m}{%
(a\left| \vec{x}\right| )^{n-3}\left( \left| \vec{x}\right| \omega \right)
^{2}}  \label{TISDS}
\end{equation}
and the total mass as $t\rightarrow \infty $ is 
\begin{equation}
{\frak M}_{{\rm dS/CFS}}^{\left( n\right) }=-V^{n-2}\frac{m\left( n-2\right) 
}{4\pi G_{n}}  \label{MA}
\end{equation}
again illustrating the dependence of the Casimir energy on the choice of
slicing: in the above, the Casimir energy does not appear in CAF slicings.

Similar results hold in the AdS case where $k=-1$. Under a coordinate
transformation, the metric (\ref{SADS}) becomes 
\begin{equation}
ds^{2}=g^{2}dr^{2}+a^{2}\omega ^{4/(n-3)}(-dt^{2}+t^{2}\sigma
_{ij}dx^{i}dx^{j}),  \label{sitadmet}
\end{equation}
with $a(r)=e^{-r/\ell }$, 
$g=2/\omega-1$ and 
$\omega (\tau ,\vec{x})=1-m/2[a(r)t]^{n-3}$, and an analogous calculation
using CAF slicing yields \ 
\begin{equation}
{\frak M}_{{\rm AdS/CFS}}^{\left( n,k=-1\right) }=V^{n-2}\frac{m\left(
n-2\right) }{4\pi G_{n}}  \label{MAdS}
\end{equation}
for the total mass as $r\rightarrow \infty $. \ 

\bigskip

\section{AdS/CFT and the brane-world}

The goal of this section is to see where the Casimir energy appears in the
context of an AdS/CFT interpretation \cite{Misao,Lisa2,Ida} of the
brane-world scenario \cite{Lisa}. Our argument is based on ref. \cite{Ida}
where the cosmological constant on the brane is set to zero. We first extend
the previous formulation\cite{Ida} to that on curved branes with non-zero
cosmological constant (See \cite{Padilla,Youm,Med} for the deSitter brane
case). Then we might expect that we can see curvature or topological effects
as the CFT contribution dual the adS bulk. In this section we consider only
five dimensions.

Recall that the following gravitation equation holds on the brane\cite{SMS}: 
\begin{equation}
{}^{(4)}G_{\mu \nu }=-\Lambda _{4}q_{\mu \nu }+8\pi G_{4}{\sf T}_{\mu \nu
}+\kappa _{5}^{4}\pi _{\mu \nu }-E_{\mu \nu },  \label{branegrav1}
\end{equation}
where $q_{\mu \nu }$ is the induced metric on the brane, 
$8\pi G_{4}:=\kappa _{5}^{4}\lambda/6 $ , 
\begin{eqnarray}
\Lambda _{4} &:&=\frac{1}{12}\left[(\kappa _{5}^{2}\lambda )^{2}-\left( 
\frac{6}{\ell }\right) ^{2}\right]  \nonumber \\
\pi _{\mu \nu } &:&=-\frac{1}{4}{\sf T}_{\mu }^{\alpha }{\sf T}_{\alpha \nu
}+\frac{1}{12}{\sf TT}_{\mu \nu }+\frac{1}{8}q_{\mu \nu }{\sf T}_{\alpha
\beta }{\sf T}^{\alpha \beta }-\frac{1}{24}q_{\mu \nu }{\sf T}^2  \nonumber
\end{eqnarray}
and 
\begin{equation}
E_{\mu \nu }:={}^{(5)}C_{\mu \alpha \nu \beta }n^{\alpha }n^{\beta }.
\label{CO2}
\end{equation}
In the above $\lambda $ is the brane tension and ${\sf T}_{\mu \nu }$ is the
energy-momentum tensor on the brane. This equation is exact if we assume
that the five dimensional Einstein equation holds and $Z_2$-symmetry. Since
only $E_{\mu \nu }$ contains bulk information, it expresses the contribution
from the Kaluza-Klein modes\cite{SMS}. Moreover $E_{\mu \nu }$ can be shown
to be identical to the CFT energy-momentum tensor as seen below.

On the other hand, from the perspective of the AdS/CFT correspondence, we
have the effective equation on the brane 
\begin{eqnarray}
{}^{(4)}G_{\mu \nu } &=&-\frac{1}{\ell }\left( \kappa _{5}^{2}\lambda -\frac{%
6}{\ell }\right) q_{\mu \nu }+8\pi G_{5}\ell ^{-1}\Bigl({\sf T}_{\mu \nu
}+T_{\mu \nu }^{({\rm CFT})}\Bigr)  \nonumber \\
&&+O({\sf T}^{2})  \label{branegrav2}
\end{eqnarray}
where $T_{\mu \nu }^{({\rm CFT)}}$ is the CFT\ energy-momentum tensor on the
brane. See Refs. \cite{Misao,Lisa2,Ida} for $\kappa _{5}^{2}\lambda =6/\ell $
cases where the cosmological constant term on the brane vanishes. Note that $%
\kappa _{5}^{2}\lambda =6/\ell$ holds exactly in Randall-Sundrum toy models
to realize the Minkowski branes.

To linear order in ${\sf T}_{\mu \nu }$ and $\left( \kappa _{5}^{2}\lambda
-6/\ell \right) $, we can see that eqs. (\ref{branegrav1}) and (\ref
{branegrav2}) are the same provided we identify the CFT energy-momentum
tensor with the bulk part of the electric Weyl tensor: 
%
\begin{equation}
8\pi G_{5}\ell ^{-1}T_{\mu \nu }^{{\rm CFT}}\simeq -E_{\mu \nu }.
\label{CO3}
\end{equation}
%
%
%
where we approximate $G_{4}$ by $G_{5}\ell ^{-1}$ and $\Lambda _{4}$ in Eq. (%
\ref{branegrav1}) by 
\begin{eqnarray}
\Lambda _{4} &=&\frac{1}{\ell }\biggl(\kappa _{5}^{2}\lambda -\frac{6}{\ell }%
\biggr)+\frac{1}{12}\biggl(\kappa _{5}^{2}\lambda -\frac{6}{\ell }\biggr)^{2}
\nonumber \\
&\simeq &\frac{1}{\ell }\biggl(\kappa _{5}^{2}\lambda -\frac{6}{\ell }\biggr)%
,
\end{eqnarray}
respectively.

Consequently we recover the AdS/CFT interpretation to leading order in $%
\left( \kappa _{5}^{2}\lambda -6/\ell \right) $ for the brane-world, but we
cannot retain this interpretation to all orders \footnote{%
It has been shown that $\pi _{\mu }^{\mu }$ is related to the conformal
anomaly\cite{Ida}}

Finally, we consider the Casimir energy. As an example let us take the
homogeneous and isotropic universe on the brane\cite{cosmos}: 
\begin{equation}
\Bigl(\frac{\partial _{\tau }a}{a}\Bigr)^{2}+\frac{k}{a^{2}}=\frac{8\pi G_{4}%
}{3}\rho +\frac{\Lambda _{4}}{3}+\frac{\kappa _{5}^{4}}{36}\rho ^{2}+\frac{%
\mu }{a^{4}},  \label{IU}
\end{equation}
where $a(\tau )$ is the scale factor, $\tau $ is the proper time on the
brane and $\mu $ is the mass parameter of the five dimensional
Schwarzshild-AdS spacetime. The last term in the right-hand side is the dark
radiation \cite{SMS,cosmos,Vincent} which comes from $E_{\mu \nu }$. The
curvature term is usual one which appears in a $4$-dimensional cosmology,
and so it is not directly related to the CFT in the brane world. This result
is differs from the viewpoint argued by Verlinde \cite{linde} because this
term is supposed to correspond to the Casimir energy of CFT on the brane.
But according to the standard AdS/CFT viewpoint such terms should arise from 
$E_{\mu \nu }$.

Although the Casimir energy does not appear in the {\it homogeneous and
isotropic} brane universe according to the standard AdS/CFT, this does not
mean that $E_{\mu \nu }$ does not contain the Casimir energy. A
consideration of fluctuations around the homogeneous and isotropic brane
universe with non-trivial topology indicates that the Casimir energy of the
fluctuations appears in $E_{\mu \nu }$ and $T_{\mu \nu }^{({\rm CFT})}$ as
usual in curved spacetimes\cite{QFT}.

\section{Summary}

In this paper, we have investigated the slicing dependence of the Casimir
energy in the context of the (A)dS/CFT correspondence. We find a significant
dependence of the Casimir energy on both the geometry and the topology of
the boundary slicing. For asymptotically flat slicings (i.e. those which
become sufficiently conformally flat), the CFT energy measured at the
boundary is same as the gravitational energy and there is no Casimir
contribution. These features are consistent with curved-space(time) quantum
field theory.

We also have discussed the AdS/CFT interpretation in the brane-world. We
could show that the bulk Weyl tensor corresponds to the CFT stress tensor.
The Casimir energy does not emerge for the observers on the brane when we
think of the homogeneous-isotropic universe on the brane. So, following the
standard AdS/CFT correspondence, it is difficult to identify the curvature
term with the Casimir energy as Verlinde did\cite{linde}. Evidently
Verlinde's argument goes beyond the scope of the standard AdS/CFT
correspondence.

The relationship between the slicing dependence of the Casimir energy in
terms of the state of the CFT \cite{State} remains an interesting subject
for future investigation.

\section*{Acknowledgments}

DI and TS would like to thank S. Hayakawa, H. Ochiai, S. Ogushi, Y. Shimizu
and T. Torii for their discussion. TS's work is partially supported by
Yamada Science Foundation. This work was supported in part by the Natural
Sciences and Engineering Research Council of Canada.

\end{document}